\documentclass[epj]{webofc}
\usepackage[utf8]{inputenc}
\usepackage[varg]{txfonts}   
\usepackage{booktabs}
\usepackage{xcolor,amsmath}
\DeclareMathOperator{\Tr}{Tr}
\DeclareMathOperator{\SU}{SU}
\definecolor{darkred}{rgb}{0.4,0.0,0.0}
\definecolor{darkgreen}{rgb}{0.0,0.4,0.0}
\definecolor{darkblue}{rgb}{0.0,0.0,0.4}
\usepackage[bookmarks,linktocpage,colorlinks,
    linkcolor = darkred,
    urlcolor  = darkblue,
    citecolor = darkgreen]{hyperref}
\usepackage{subcaption}
\wocname{EPJ Web of Conferences}
\woctitle{Lattice2017}
\begin{document}
%
\selectlanguage{english}
\title{%
CLS 2+1 flavor simulations at physical light- and strange-quark masses
}
\author{%
\firstname{Daniel} \lastname{Mohler}\inst{1,2}\fnsep\thanks{Speaker, \email{damohler@uni-mainz.de}} \and
\firstname{Stefan}  \lastname{Schaefer}\inst{3} \and
\firstname{Jakob} \lastname{Simeth}\inst{4}
}
\institute{%
Helmholtz-Institut Mainz, 55099 Mainz, Germany
\and
Johannes Gutenberg Universit\"at Mainz, 55099 Mainz, Germany
\and
John von Neumann Institute for Computing (NIC), DESY, Platanenallee 6, 15738 Zeuthen, Germany
\and
Institute for Theoretical Physics, University of Regensburg, 93040 Regensburg, Germany
}
\abstract{%
We report recent efforts by CLS to generate an ensemble with physical light- and strange-quark masses in a lattice volume of $192\times96^3$ at $\beta=3.55$ corresponding to a lattice spacing of $0.064$~fm. This ensemble is being generated as part of the CLS 2+1 flavor effort with improved Wilson fermions. Our simulations currently cover 5 lattice spacings ranging from $0.039$~fm to $0.086$~fm at various pion masses along chiral trajectories with either the sum of the quark masses kept fixed, or with the strange-quark mass at the physical value. The current status of simulations is briefly reviewed, including a short discussion of measured autocorrelation times and of the main features of the simulations. We then proceed to discuss the thermalization strategy employed for the generation of the physical quark-mass ensemble and present first results for some simple observables. Challenges encountered in the simulation are highlighted.
}
\maketitle
\section{Introduction}\label{sec:introduction}

The CLS (Coordinated Lattice Simulations) consortium with members in Denmark,
Germany, Italy, Spain, and Switzerland has embarked on a program to generate
2+1 flavor gauge ensembles with $O(a)$ improved Wilson fermions and a
L\"uscher-Weisz gauge action with tree-level coefficients. A special feature
is the use of open boundary conditions in the time direction (to avoid
topological freezing at fine lattice spacings \cite{Luscher:2011kk}) for the bulk of
ensembles generated to date. Another feature is the introduction of a small
twisted-mass term for the light (up and down) quarks in the simulation together with subsequent reweighting to
zero twisted mass (twisted mass reweighting) \cite{Luscher:2012av}, to avoid
accidental near-zero modes of the lattice Dirac operator. Most of the ensembles generated
initially \cite{Bruno:2014jqa} are along a trajectory keeping the trace of the bare
quark-mass matrix fixed, $\Tr(M)=\mathrm{const}$. More recently, these
ensembles have been supplemented by ensembles along a trajectory with fixed
strange quark mass $m_s\approx m_{s,phys}$ \cite{Bali:2016umi} and with
$m_s=m_l$ at different values of $\Tr(M)$.

For the simulation, the highly flexible \texttt{openQCD} package \cite{OpenQCD} is used. Among its main features are the
use of nested hierarchical integrators (for the implementation see
\cite{Luscher:2012av}), Hasenbusch-style frequency splitting \cite{Hasenbusch:2001ne} with an arbitrary number of pseudofermion pairs,
RHMC \cite{Kennedy:1998cu} (plus reweighting) for the strange quark,
deflation acceleration along the HMC trajectory \cite{Luscher:2007se,Luscher:2007es} and a chronological inverter \cite{Brower:1995vx}.
The package implements a number of solvers.

\begin{figure}[tbh]
\centering
\sidecaption
\includegraphics[clip,width=0.48\textwidth]{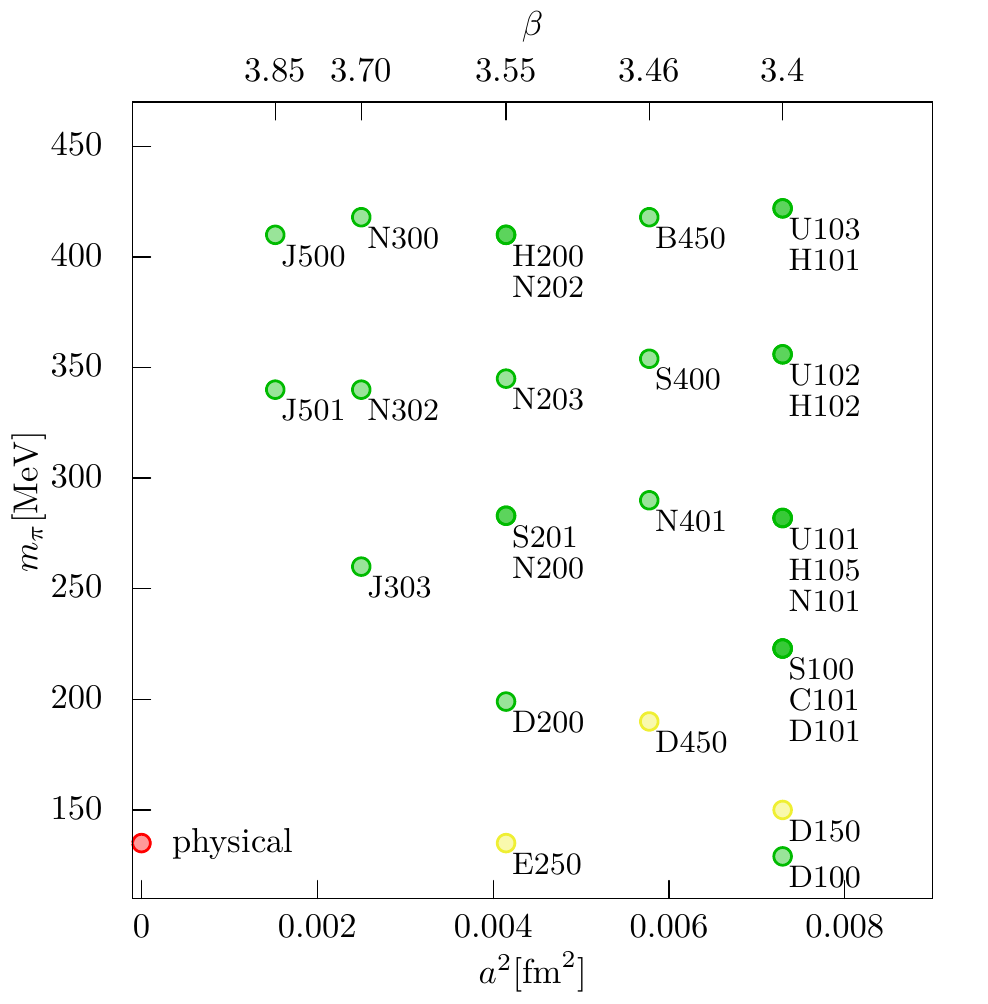}
\includegraphics[clip,width=0.48\textwidth]{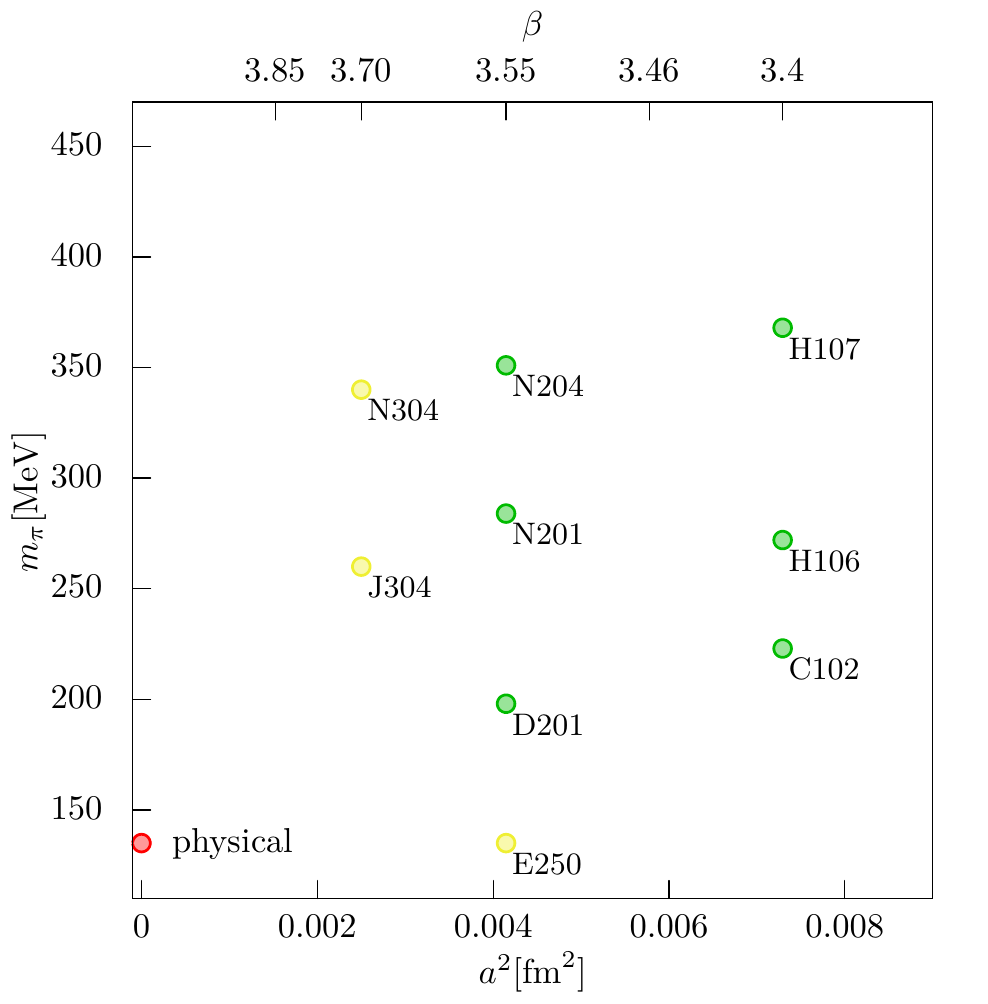}
\caption{Landscape of CLS 2+1-flavor ensembles with $\Tr(M)=\mathrm{const}$ (left pane)
  and $m_s=\mathrm{const}$ (right pane). In both cases the lattice spacing squared is
  displayed on the x-axis, while the y axis shows the pion mass. Ensembles
  still in production are shown in yellow, while ensembles considered complete
are shown in green. The physical light-quark mass ensemble E250 discussed in these
proceedings is shown in both plots. Multiple ensemble names next to a single
dot indicate ensembles with different volumes at the (otherwise) same set of simulation parameters.}
\label{ensembles1}
\end{figure}

\begin{figure}[tbh]
\centering
\includegraphics[clip,width=0.48\textwidth]{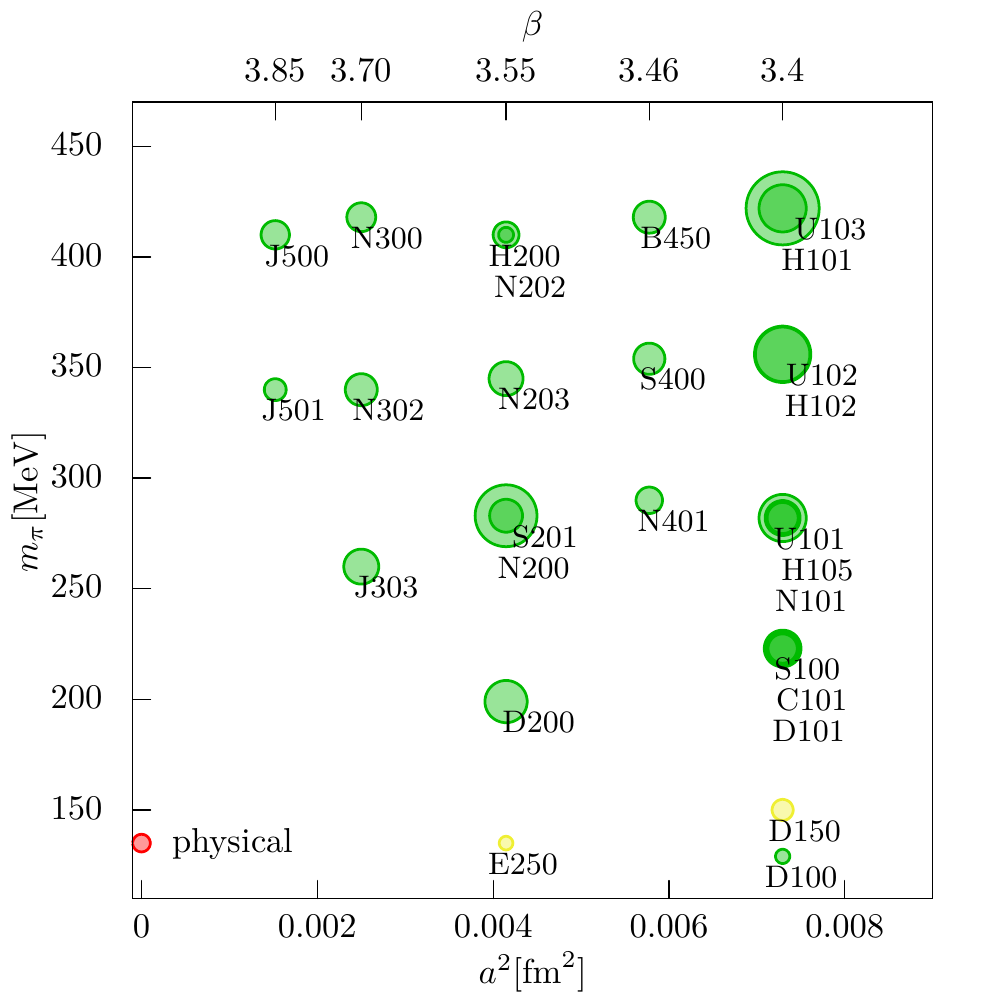}
\includegraphics[clip,width=0.48\textwidth]{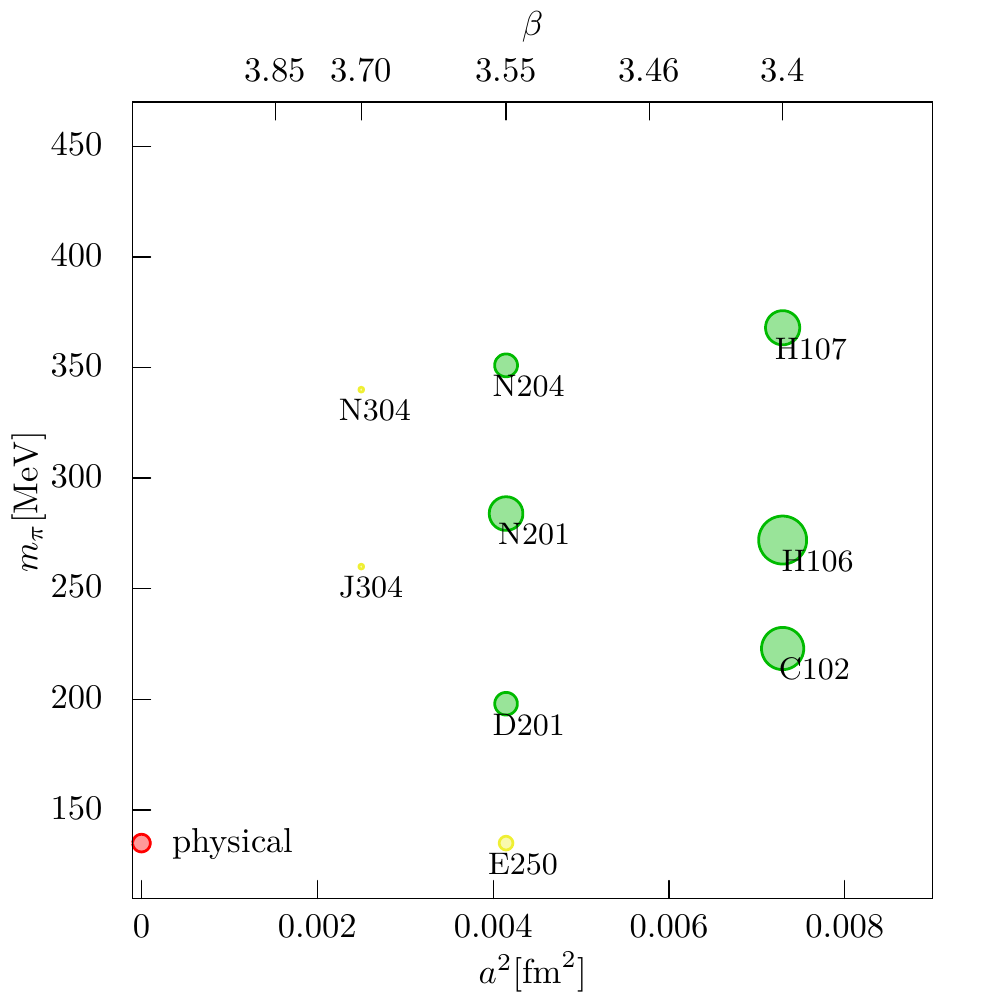}
\caption{Landscape of CLS 2+1-flavor ensembles with $\Tr(M)=\mathrm{const}$ (left pane)
  and $m_s=\mathrm{const}$ (right pane). The area of the circles in this plot is
  proportional to the number of MDU divided by the
  largest integrated autocorrelation time $\tau_{int}$}
\label{ensembles2}
\end{figure}

Figure \ref{ensembles1} shows the current set of ensembles for both
$\Tr(M)=\mathrm{const}$ (left pane) and $m_s=\mathrm{const}$ (right pane). The ensembles are
labeled by a letter (denoting the aspect ratio $T/L$) and three numerical
digits (the first digit encodes $\beta$ and therefore the lattice spacing).
Our current library features ensembles at 5 lattice spacings (ranging from
$0.039$~fm to $0.086$~fm) and with a range of pion masses $M_\pi\le 420$~MeV. For some sets
of parameters, multiple lattice volumes exist, enabling us to control finite volume effects. Figure \ref{ensembles2} shows the
same set of ensembles, now highlighting the current set of statistics. We
typically generate chains of roughly 4000 molecular dynamics units (MDU), and
save a gauge configuration every 4 MDU. The choice of target statistics is made
considering the largest integrated autocorrelation time $\tau_{int}$ (often
given by the Yang Mills action density at finite flow time). 

\begin{figure}[tbh]
\centering
\includegraphics[clip,width=0.48\textwidth]{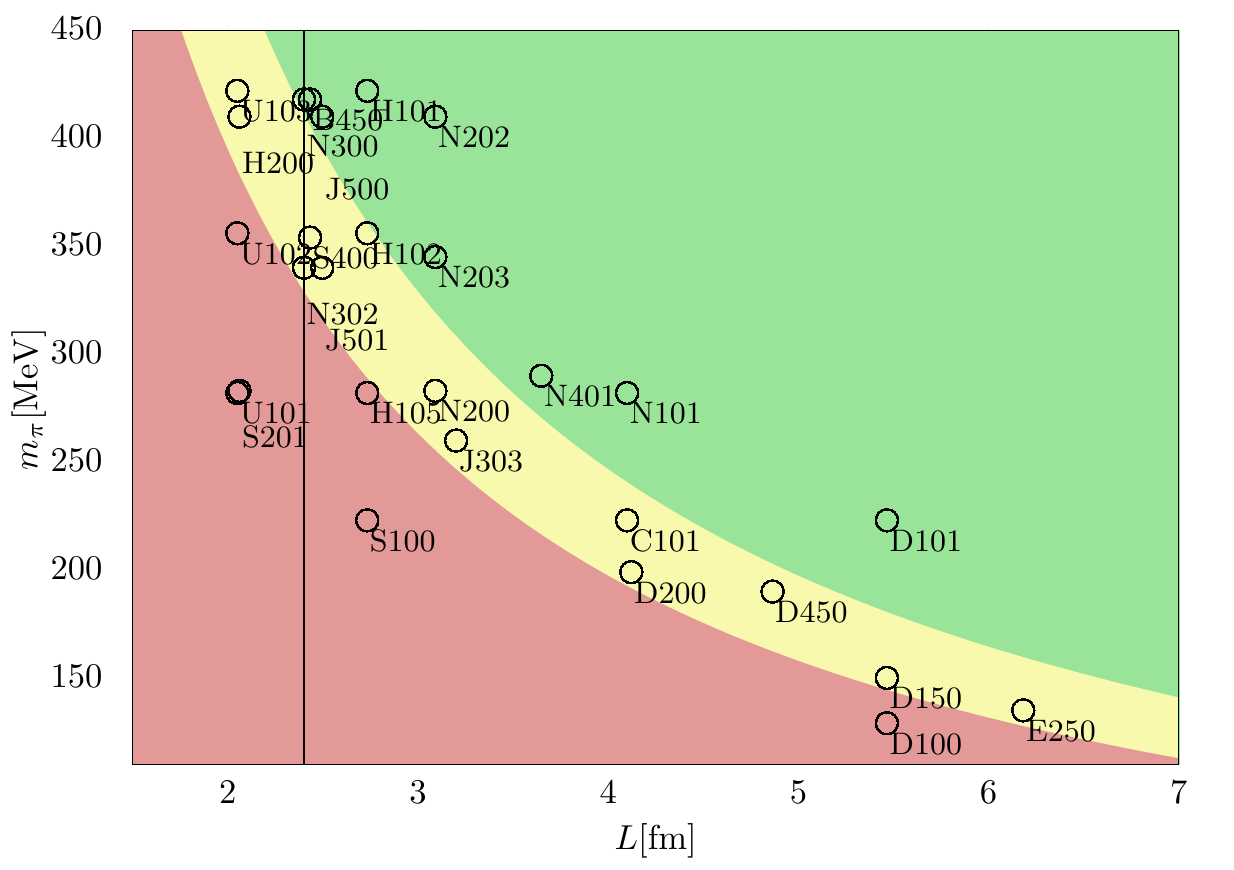}
\includegraphics[clip,width=0.48\textwidth]{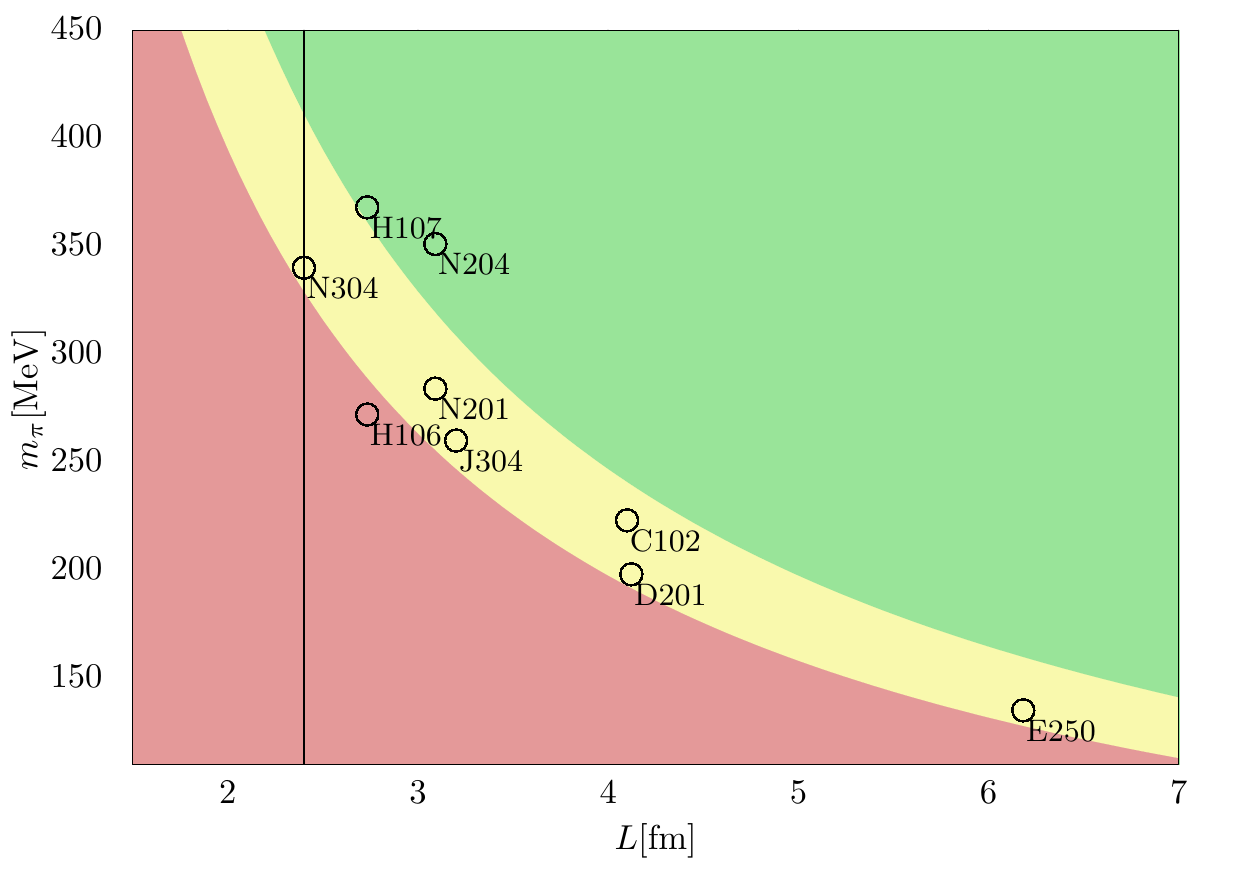}
\caption{Landscape of CLS 2+1-flavor ensembles with $\Tr(M)=\mathrm{const}$ (left pane)
  and $m_s=\mathrm{const}$ (right pane). The spatial extent $L$ of the ensemble is
  used as the x-axis value. The region of the plot with  $m_\pi L\le4$ is
  colored red, the region with $4 \le m_\pi L\le5$ is colored yellow and
  the region with $5\le m_\pi L$ is colored green.}
\label{ensembles3}
\end{figure}

Finally, Figure \ref{ensembles3} shows the spatial extent $L$ of the
ensembles. Most production ensembles feature $m_\pi L\ge 4$, ensuring that
exponentially suppressed volume effects are small. For some parameter sets,
smaller volumes to check for and control finite size effects have also been generated.


\section{Autocorrelation times towards the physical point}

In this section we will provide a brief update on estimated autocorrelation
times for the Yang-Mills action density at flow time $t_0$ determined by the
condition $t^2\langle E\rangle=0.3$ \cite{Luscher:2010iy}. While the open boundary conditions in time
\cite{Luscher:2011kk} avoid topological freezing at fine lattice spacing $a$, it is expected that the autocorrelation
time in this observable increases significantly as the lattice spacing is
decreased. In a previous global fit \cite{Bruno:2014jqa} of data from the
initial set of 2+1 flavor CLS ensembles, the autocorrelation time was determined to be well described by  $\tau_{\mathrm{exp}}=14(3)\frac{t_0}{a^2}$.

\begin{figure}[tbh]
\centering
\begin{subfigure}[c]{0.3\textwidth}
\subcaption{\textbf{N203}\\$\tau_{\mathrm{int}}=60(32)$ MDU}
\includegraphics[clip,width=1.0\textwidth]{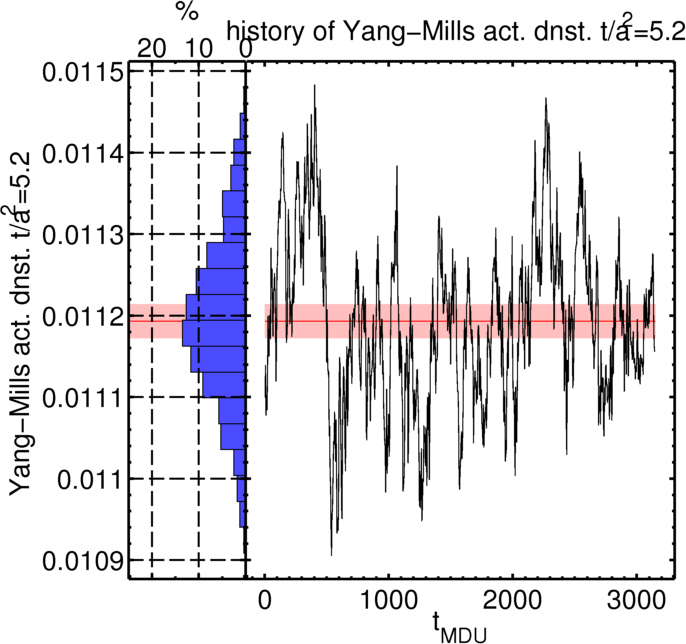}
\end{subfigure}
\begin{subfigure}[c]{0.3\textwidth}
\subcaption{\textbf{N302}\\$\tau_{\mathrm{int}}=49(22)$ MDU}
\includegraphics[clip,width=1.0\textwidth]{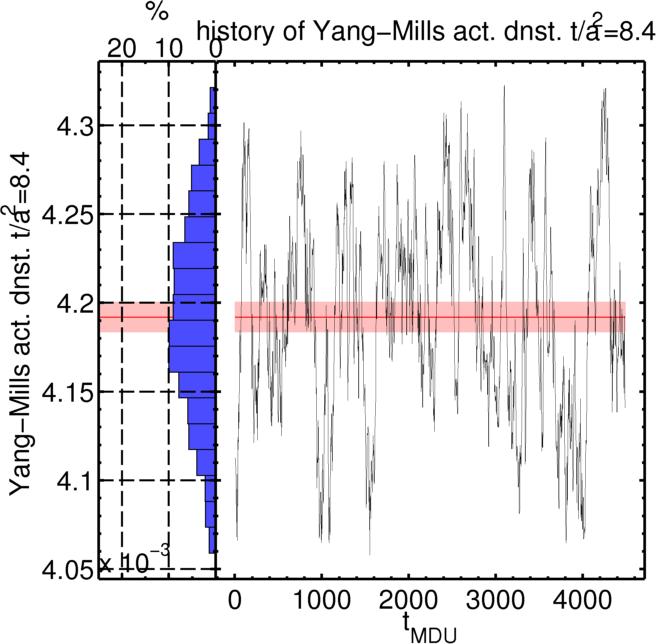}
\end{subfigure}
\begin{subfigure}[c]{0.3\textwidth}
\subcaption{\textbf{J501}\\$\tau_{\mathrm{int}}=150(98)$ MDU}
\includegraphics[clip,width=1.0\textwidth]{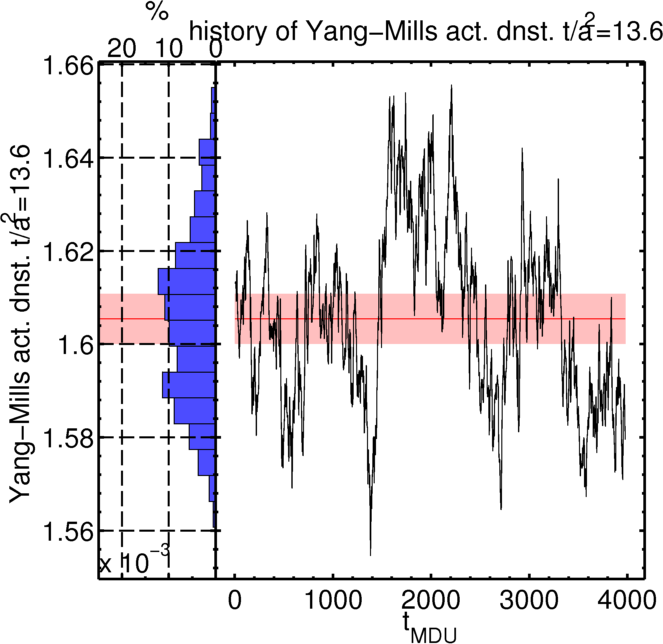}
\end{subfigure}
\caption{History of the YM action density  at flow  time $t_0$ along the MD
  trajectory.The three panes show results for $a\approx0.064$fm (a),
  $a\approx0.050$fm (b), and $a\approx0.039$fm (c).}
\label{autocorrelation1}
\end{figure}

\begin{figure}[tbh]
\centering
\begin{subfigure}[c]{0.3\textwidth}
\subcaption{\textbf{N203}\\$\tau_{\mathrm{int}}=60(32)$ MDU}
\includegraphics[clip,width=1.0\textwidth]{N203_Yt-hist.png}
\end{subfigure}
\begin{subfigure}[c]{0.3\textwidth}
\subcaption{\textbf{N200}\\$\tau_{\mathrm{int}}=45(21)$ MDU}
\includegraphics[clip,width=1.0\textwidth]{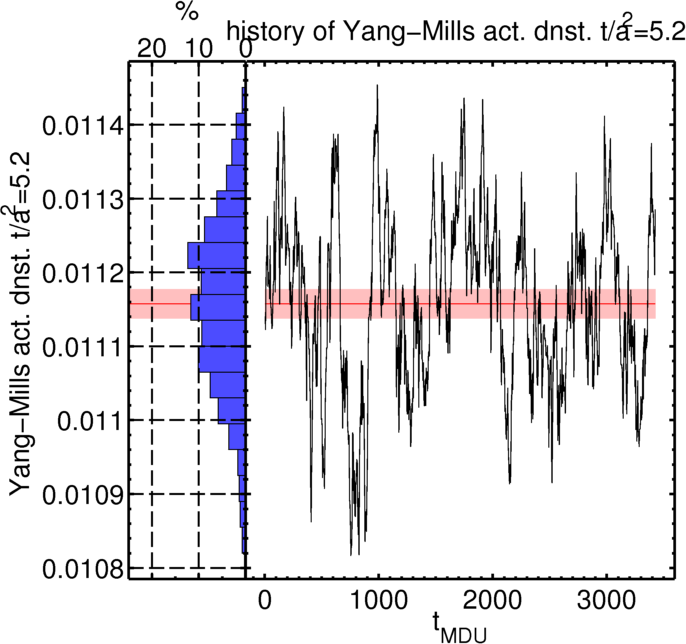}
\end{subfigure}
\begin{subfigure}[c]{0.3\textwidth}
\subcaption{\textbf{D200}\\$\tau_{\mathrm{int}}=101(55)$ MDU}
\includegraphics[clip,width=1.0\textwidth]{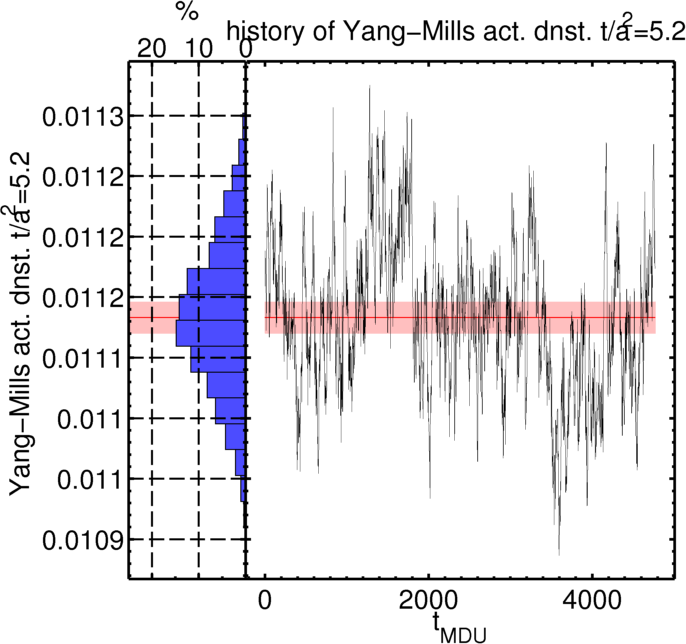}
\end{subfigure}
\caption{History of the YM action density  at flow  time $t_0$ along the MD
  trajectory.The three panes show results at fixed
  lattice spacing $a\approx0.064$fm for $m_\pi\approx340$~MeV (a),
  $m_\pi\approx280$~MeV (b), and $m_\pi\approx200$~MeV (c).}
\label{autocorrelation2}
\end{figure}

Figure \ref{autocorrelation1} shows the MD history of the YM action density at
flow  time $t_0$ for fixed pion mass $m_\pi\approx 340$~MeV and for three
lattice spacings. The corresponding integrated autocorrelation time
$\tau_{\mathrm{int}}$ is given above each subfigure. Even with chains longer than
3000 MDU the statistical uncertainty on $\tau_{int}$ is large. The current results are
consistent with the expected increase of the autocorrelation time
\cite{Bruno:2014jqa}. Figure \ref{autocorrelation2} shows the MD history of
the YM action density at flow time $t_0$ for three different pion masses along
the trajectory with $\Tr(M)=\mathrm{const}$. Unlike in simulations by the MILC collaboration at fixed strange-quark mass \cite{Bernard:2017npd}, no clear pattern is
seen with our current statistics. 

\section{Thermalization of a physical light-quark mass ensemble}

While the current CLS 2+1 flavor ensembles cover a wide range of parameters,
the accurate determination of some quantities of phenomenological interest
are currently limited by our control of systematic uncertainties associated
with the chiral extrapolation. An important example is the calculations of the
hadronic contributions to the anomalous magnetic moment of the muon
(see \cite{DellaMorte:2017dyu} for a CLS based determination). Beyond specific observables, our determination of the lattice scale
\cite{Bruno:2016plf} would also greatly benefit from simulations with physical light- and
strange-quark masses. 

In these proceedings we report on our efforts to thermalize an ensemble (E250)
at physical light- (mass degenerate up and down) and strange-quark mass and
with $\beta=3.55$ corresponding to a lattice spacing $a\approx0.064$~fm. At
this lattice spacing, the exponential autocorrelation time $\tau_{exp}$  is
not yet completely dominated by topology, and therefore periodic boundary
conditions are chosen. To keep $m_\pi L\ge 4$, we choose a lattice volume
$T\times L^3=192\times 96^3 a^4$. We work with the following thermalization
strategy:
\begin{itemize}
\item Start from an $\SU(3)$ run with $T\times L^3=64\times 48^3 a^4$, 3 mass-degenerate light
  quarks and periodic boundary conditions generated for non-perturbative
  renormalization. Perform a number of runs (updating the light- and
  strange-quark hopping parameters to their target values ) and thermalize this small volume.
\item Triple the time extent and partially thermalize this intermediary
  ensemble of size $192\times 48^3 a^4$.
\item Double the spatial extent and thermalize the resulting ensemble of
  size $192\times 96^3 a^4$.
\end{itemize}

\subsection{Status of gauge field generation for E250}

\begin{figure}[tbh]
\centering
\includegraphics[clip,width=0.32\textwidth]{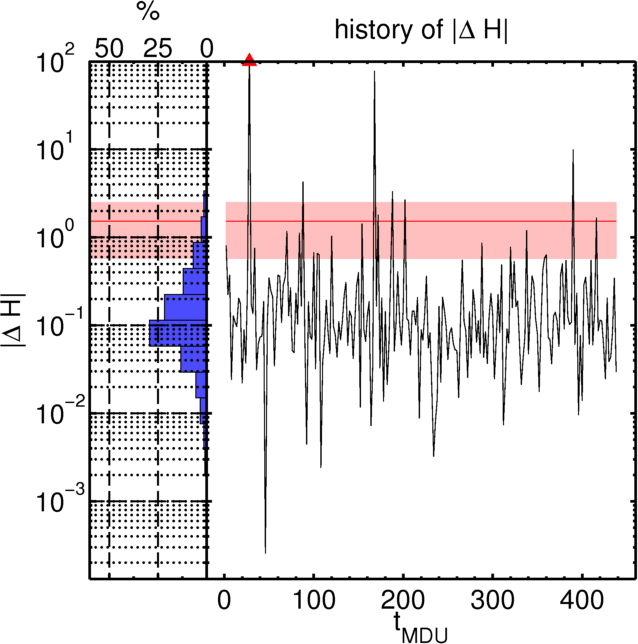}
\includegraphics[clip,width=0.32\textwidth]{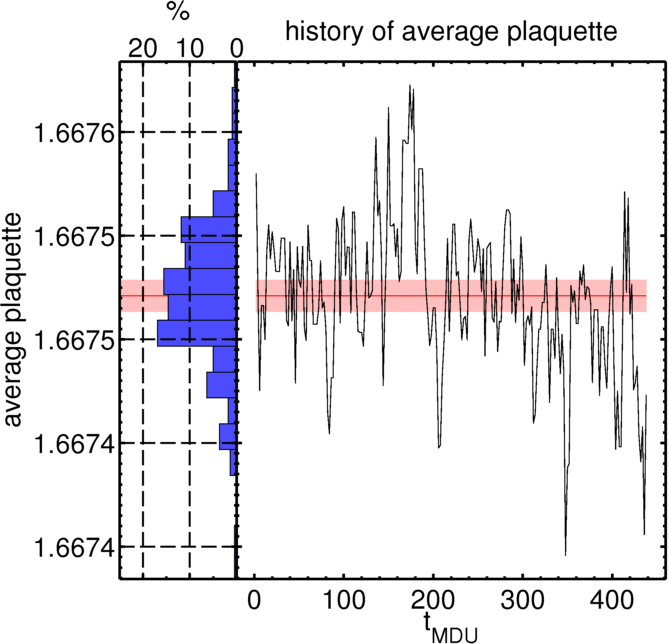}
\includegraphics[clip,width=0.32\textwidth]{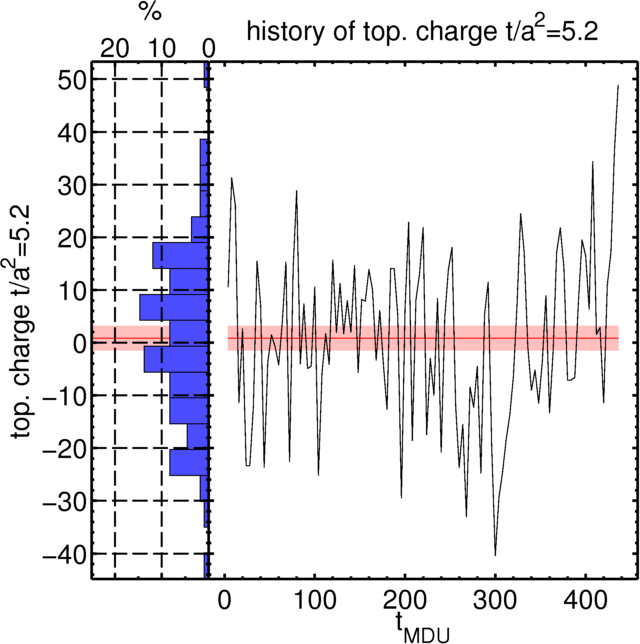}
\caption{Left pane: MD history of $\Delta H$ for the physical mass run E250. Mid pane: MD history of the
  average plaquette. Right pane: History of the topological charge at the
  center of the lattice.}
\label{e250:plots}
\end{figure}

The thermalization runs on the small and intermediate volume have been
successfully completed and we are currently producing a first production chain.
At the time of Lattice2017, there was a chain of 436 MDU, corresponding to 109
saved configurations, some of which might not be fully thermalized. The measured acceptance is $0.872(27)$. Figure
\ref{e250:plots} shows plots for the history of the hamiltonian deficit $\Delta H$ seen in the
Monte-Carlo accept/reject step, for the average plaquette, and for the
topological charge at the center of the lattice. While a proper analysis will need a much
longer Monte-Carlo chain, the right panel of of Figure \ref{e250:plots}
suggests that at this lattice spacing, the slowing down of topological tunneling is not yet a
serious issue.

\begin{figure}[tbh]
\centering
\includegraphics[clip,width=0.48\textwidth]{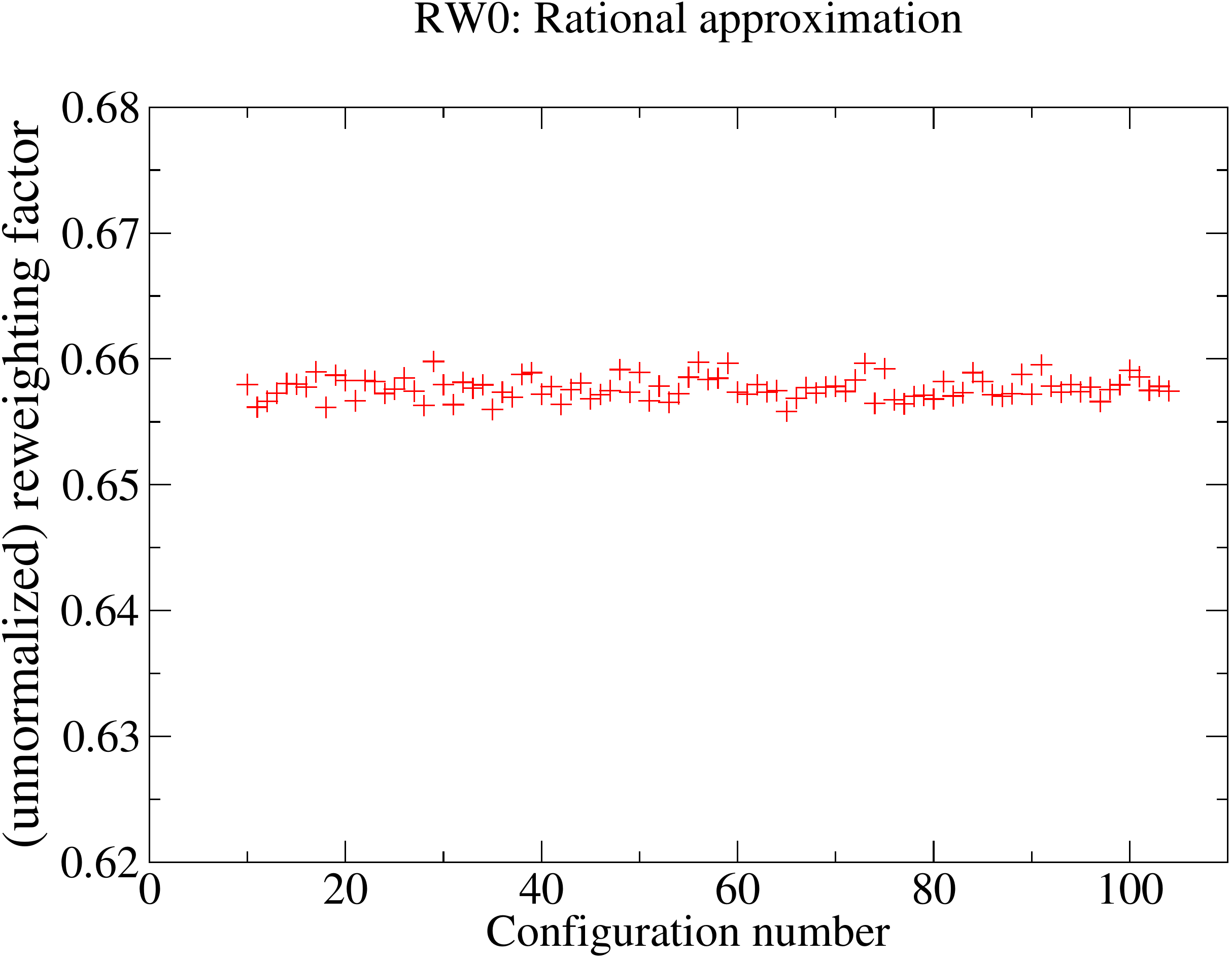}
\includegraphics[clip,width=0.48\textwidth]{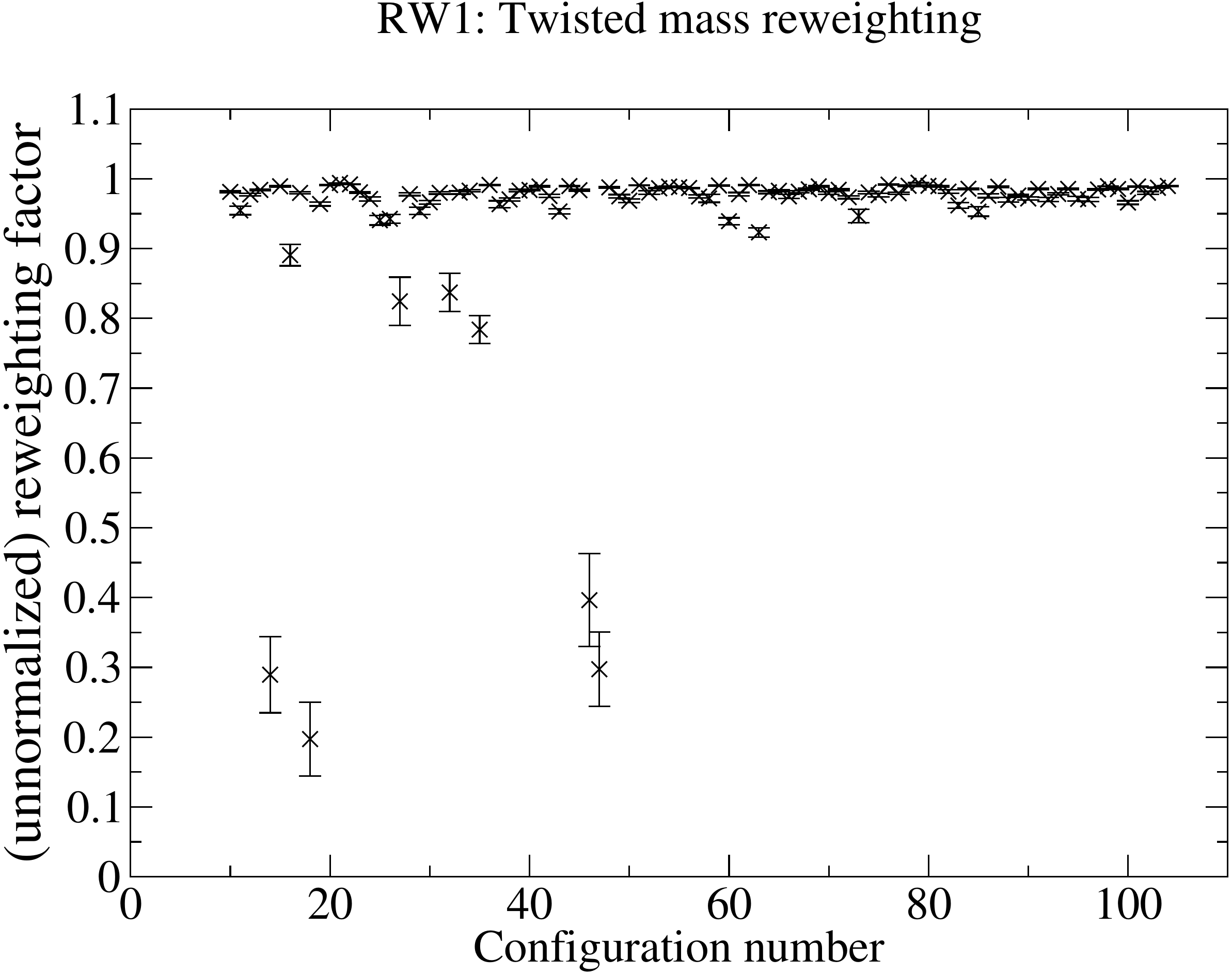}
\caption{Reweighting factors for a subset of the configuration produced of
  E250. Fluctuations in the reweighting factor associated with the rational
  approximation are very mild (left pane; red). The reweighting factor for the
twisted mass reweighting (right pane; black) shows larger fluctuations.}
\label{e250:reweighting}
\end{figure}

\begin{figure}[tbh]
\centering
\includegraphics[clip,width=0.48\textwidth]{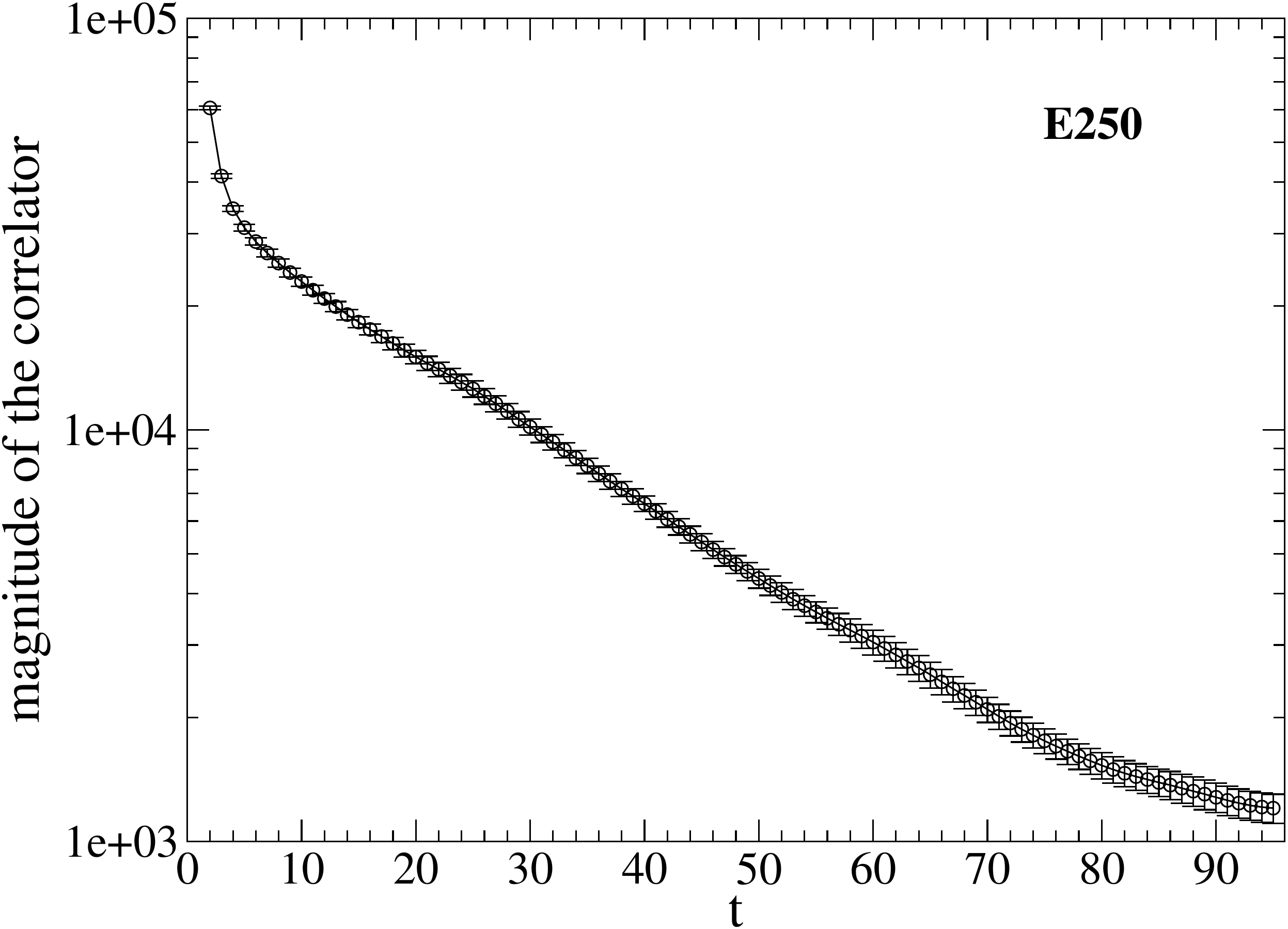}
\includegraphics[clip,width=0.48\textwidth]{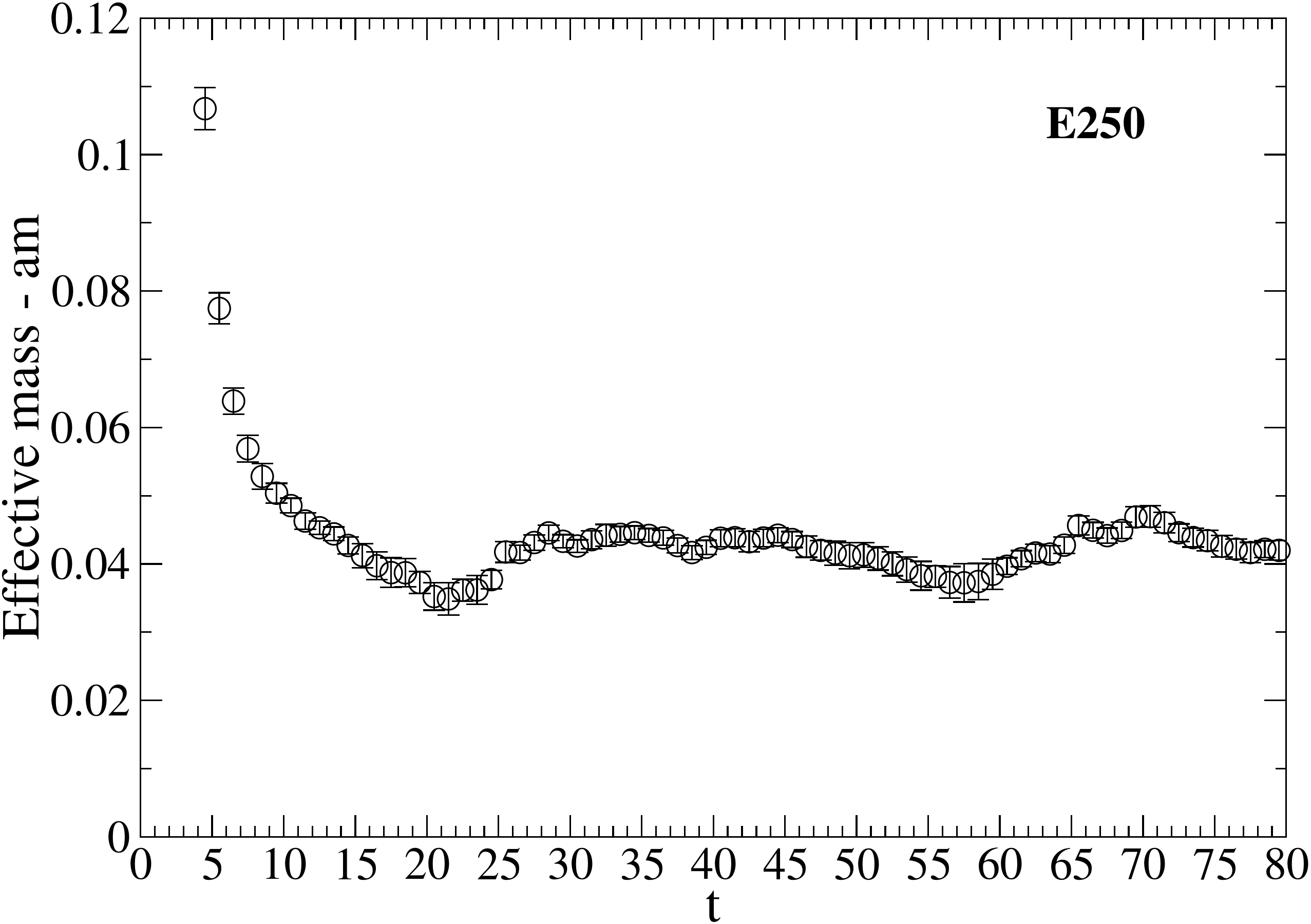}
\caption{Left pane: Pion correlator from 70 configurations of E250. Right
  pane: Corresponding effective mass. For details of the correlator
  construction please refer to the text.}
\label{e250:pion}
\end{figure}

Figure \ref{e250:reweighting} shows the reweighting factors associated with
the rational approximation for the strange quark and for the reweighting to
zero twisted mass. The reweighting factor for the twisted mass reweighting is
estimated using 24 random sources and no intermediate value for the twisted
mass parameter $\mu$. For the full ensemble, this estimate might have to be
improved, as occasional small reweighting factors are not estimated very
accurately. \footnote{The fluctuations of the twisted mass reweighting factor are also
  somewhat larger than desirable and a slightly smaller twisted mass parameter
  (chosen as $\mu=0.0001$ for the current run) might have been a better choice.}

 Figure \ref{e250:pion} shows results for the pion correlator and pion
effective mass using random wall sources on the same time slice for 70
configurations. With this procedure, autocorrelations in the pion correlator
are clearly visible. For future measurements we will make use of the large
time extent of the lattice and the periodic boundary conditions by randomly
shifting source locations. The current results indicate $m_\pi\approx
130(3)$~MeV, reasonably close to the physical pion in the isospin limit
 and without electromagnetic contributions \cite{Aoki:2013ldr}, with a mass of $134.8(3)$~MeV.

\subsection{Challenges encountered during the thermalization}

The initial thermalization runs with $T\times L^3=64\times 48^3 a^4$ proceeded
smoothly. The same was the case for runs at intermediate volume, which needed
various minor parameter adjustments, such as more frequent updates of the
deflation subspace along the MD trajectory. Switching to the large volume
with $T\times L^3=192\times96^3 a^4$, which was an unprecedented lattice size for
CLS simulations using the openQCD Software, a change to a large deflation blocksize was needed in order to maintain a manageable
  size of the little Dirac operator. While the deflation subspace for
  CLS runs in smaller volumes is typically of order $4^3\times 8$ we had to
  use a much larger blocksize ($8\times 4\times 8^2$) to achieve a stable
  run. For this setup the resulting iteration counts are somewhat higher than
  desirable, as deflation is not as efficient. This indicates that a multigrid setup with 3 levels might be preferable for this
  lattice volume\footnote{It is currently not clear that this would pay off
    for E250, as possible run parameters would also be quite limited}. Such a
  setup is currently not possible with openQCD. For even larger lattices
  further algorithmic improvements or a paradigm shift on how lattice QCD simulations are
  carried out \cite{Luscher:2017cjh} would be needed.

\section{Conclusions and Outlook}\label{sec:conclusions}

The CLS consortium continues its effort to produce a large library of
high-quality 2+1 flavor gauge configurations \cite{Bruno:2014jqa,Bali:2016umi}. As part of this effort
we started production of an ensemble with lattice spacing $a\approx 0.064$~fm and
(close-to) physical light- and strange-quark masses. This ensemble will play a
crucial role in reducing systematic uncertainties for ongoing projects within
the CLS consortium. Examples of such projects (with a focus on activities by the Mainz group) include
calculation of the hadronic contribution to $a_\mu$
\cite{Wittig:2017lat,Gerardin:2017lat}, ongoing baryon structure calculations
\cite{Djukanovic:2016ocj}, a program on lattice spectroscopy and scattering \cite{Hoerz:2017lat}, and high precision
scale setting \cite{Bruno:2016plf}. While no significant issues were encountered during
thermalization, simulations with a large lattice volume, such as E250, might
profit from a 3~level multigrid setup not currently available within
the openQCD package.


\section*{Acknowledgements}

Small and intermediary volume runs were performed on the BlueGene Q ``JUQUEEN'' at Forschungszentrum J\"ulich as
part of Gauss project HMZ21. The large volume runs were performed on the MogonII
cluster at Johannes Gutenberg-Universit\"at Mainz. We thank the staff of the
high-performance computing group at the ZDV, JGU Mainz as well as Dalibor
Djukanovic for their support. DM acknowledges insightful discussions with and
contributions from Tim Harris, Marco C\`e and Ben H\"orz. We thank our colleagues in CLS for the joint effort in the generation of the gauge field ensembles which form a basis for the here described computation. 

\clearpage
\bibliography{lattice2017}

\end{document}